\documentclass[acmsmall,nonacm,screen]{acmart}
\usepackage{tcolorbox}
\tcbset{
  colback=gray!10,
  colframe=black!40,
  fonttitle=\bfseries,
  boxrule=0.5pt,
  top=1pt,
  bottom=1pt,
  left=4pt,
  right=4pt,
  width=0.8\textwidth, 
}

\usepackage{tabularx}
\usepackage{booktabs}

\AtBeginDocument{%
  }

\setcopyright{acmlicensed}
\copyrightyear{2026}
\acmYear{2026}
\acmDOI{XXXXXXX.XXXXXXX}
\acmConference[Conference acronym 'XX]{Make sure to enter the correct
  conference title from your rights confirmation email}{June 03--05,
  2018}{Woodstock, NY}

\usepackage{wrapfig}
\begin{document}

\title{SmartOracle - 
An Agentic Approach to Mitigate Noise in Differential Oracles }

\author{Srinath Srinivasan}
\email{ssrini27@ncsu.edu}
\affiliation{%
  \institution{North Carolina State University}
  \city{Raleigh}
  \state{North Carolina}
  \country{USA}
}

\author{Tim Menzies}
\affiliation{%
  \institution{North Carolina State University}
  \city{Raleigh}
  \state{North Carolina}
  \country{USA}
}
\email{tjemenzie@ncsu.edu}

\author{Marcelo d'Amorim}
\affiliation{%
  \institution{North Carolina State University}
  \city{Raleigh}
  \state{North Carolina}
  \country{USA}
}
\email{damorim@ncsu.edu}
\renewcommand{\shortauthors}{Srinivasan et al.}

\begin{abstract}
Differential fuzzers detect bugs by executing identical inputs across distinct implementations of the same specification, such as JavaScript interpreters. Validating the outputs requires an oracle and for differential testing of JavaScript, these are   constructed manually, making them expensive, time-consuming, and prone to false positives. Worse, when the specification evolves, this manual effort   must be repeated.

Inspired by the success of agentic systems in other SE domains, this paper introduces \textsc{SmartOracle}. \textsc{SmartOracle} decomposes the manual triage workflow into specialized Large Language Model (LLM) sub-agents. These agents synthesize independently gathered evidence from terminal runs and targeted specification queries to reach a final verdict.

For historical benchmarks, \textsc{SmartOracle} achieves 0.84 recall with an 18\% false positive rate. Compared to a sequential Gemini 2.5 Pro baseline, it improves triage accuracy while reducing analysis time by 4$\times$ and API costs by 10$\times$.   In active fuzzing campaigns, \textsc{SmartOracle} successfully identified and reported previously unknown specification-level issues across major engines, including bugs in V8, JavaScriptCore, and GraalJS.

The success of \textsc{SmartOracle}'s agentic architecture  on Javascript suggests it might be useful other software systems- a research direction we will explore in future work.
\end{abstract}

\ccsdesc[500]{Software and its engineering~Software testing and debugging}
\keywords{Fuzzing, Agent, LLM, Oracle, JavaScript}

\received{20 February 2007}
\received[revised]{12 March 2009}
\received[accepted]{5 June 2009}

\maketitle

\section{Introduction}

Software testing remains critical to ensuring software 
reliability~\cite{swebok4}. Fuzz testing, in particular, is a 
well-established approach to uncover vulnerabilities in real-world 
systems~\cite{miller1990, zhu2022fuzzing}. However, without 
sophisticated test oracles, fuzzing is restricted to finding crashes. 
Differential oracles~\cite{mckeeman1998differential} circumvent this 
limitation by comparing the outputs of different system versions 
(e.g., compilers or runtimes) on identical inputs; discrepancies 
indicate potential \textit{functional bugs}. While adopted in several 
domains (e.g., database engines~\cite{yang2024towards}, 
compilers~\cite{yang2011finding}, runtimes, etc.), differential 
fuzzing faces two fundamental challenges:

\begin{enumerate}
    \item \textbf{Noise (False Positives):} Most output differences 
    are legitimate, stemming from unspecified behaviors, distinct 
    optimization paths, or evolving specifications. Without effective 
    filtering, differential fuzzing reports an overwhelming volume of 
    benign discrepancies, drowning developers in noise~\cite{lima2021}.
    \item \textbf{Cost:} As a consequence of this noise, the manual 
    effort required to distinguish true bugs from benign differences 
    becomes prohibitive. While rule-based oracles can mitigate this, 
    they require constant, labor-intensive maintenance to keep pace 
    with evolving software specifications~\cite{bushart2023resolfuzz}.
\end{enumerate}

\begin{figure}[t]
    \centering
    \includegraphics[width=1\linewidth]{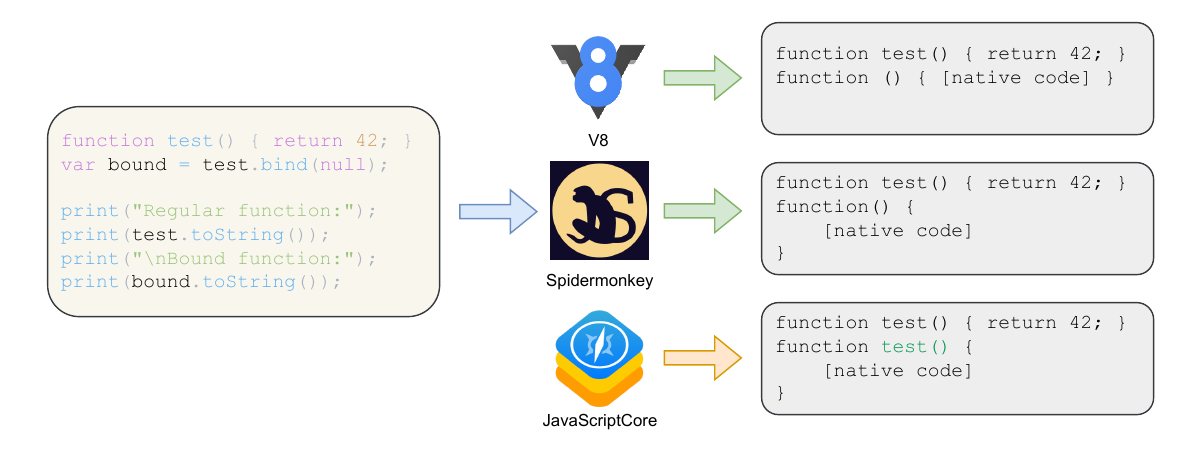}
    \caption{Example of benign output differences across three 
    JavaScript engines. While the outputs differ textually, all are 
    compliant with the ECMA-262 specification.}
    \label{fig:unspecified_behavior}
\end{figure}

Fig.~\ref{fig:unspecified_behavior} illustrates the brittleness of rule-based approaches to triage warnings from differential oracles. We focus on testing JS interpreters. The input code on the left constructs 
a bound function from a standard declaration and invokes \texttt{Function.prototype.toString()} 
to observe its string representation. As shown, 
V8~\cite{v8js} and SpiderMonkey~\cite{spidermonkey} return an 
anonymous string, while JavaScriptCore (JSC)~\cite{javascriptcore} 
preserves the function name. This divergence is explicitly sanctioned 
by ECMAScript Section 20.2.3.5~\cite{ECMA262_2025}, which permits an 
``implementation-defined'' source representation for objects lacking 
a \texttt{[[SourceText]]} slot. To resolve this, a human expert must 
isolate the code responsible for the divergence, cross-reference the 
relevant specification, and verify the interpretation through dynamic 
execution. Only then can they engineer a precise filtering rule. 
Maintaining static parsing rules for thousands of such permissible 
variations is prohibitively labor-intensive and error prone, highlighting the need 
for generalized approaches capable of reasoning about specification 
violations without relying on brittle heuristics.

A problem with all the above is that it relies too much on manual coding.
As discussed in our literature review (see Section 2.5),
for Javascript interpreters, this process is not automated.
Hence, it is time-consuming, expensive, and error-prone.
For example, while studying fuzzing Javascript, 
  recent techniques like 
AccuOracle~\cite{accuoracle}, FuzzJIT~\cite{wang2023fuzzjit}, and 
DUMPLING~\cite{wachter2025dumpling} have improved precision, they 
remain largely rule-based or tailored to specific subsystems. 

Other SE domains have explored automation with {\em agentic systems}
(communities of co-operating LLM-agents, each of which performs a sub-task).
An example of such agents in software testing is CANDOR \cite{xu2025hallucinationconsensusmultiagentllms}, a
multi-agent framework where specialized agents (Requirement Engineer, Panelist,
Interpreter, and Curator) collaborate through panel discussion to generate
accurate JUnit test oracles by consensus, achieving a 21.1 percentage point improvement.


While agentic architectures have 
seen adoption in other fields, their application to differential 
oracles for JavaScript remains unexplored. 
Hence, we propose \textsc{SmartOracle}, an 
autonomous agentic system as a differential oracle. Rather than utilizing a single monolithic 
prompt, \textsc{SmartOracle} coordinates specialized sub-agents to 
automate the analysis. A \textit{Discrepancy Finder} isolates the 
precise code segments responsible for the divergence, while a 
\textit{Specification Checker} finds the expected behavior by 
querying the relevant standards to establish ground truth without 
manual rule maintenance. Finally, a dedicated \textit{Critic} mitigates 
hallucination by verifying findings via terminal execution. This 
architecture transforms opaque logs into structured, evidence-based 
decisions. Furthermore, to enable the scalable evaluation of such 
agents, we introduce a semi-supervised labeling strategy that 
efficiently establishes ground truth for large datasets.

This work addresses three high-level research questions:

\begin{enumerate}
    \item \textit{\textbf{RQ1 (Efficacy):} Can agentic systems like \textsc{SmartOracle} 
    effectively uncover true differential testing bugs on Javascript while 
    filtering false positives?}
    \item \textit{\textbf{RQ2 (Bug-finding):} Can agentic systems like \textsc{SmartOracle} 
    discover novel bugs in the latest versions of major JavaScript 
    runtimes?}
    \item \textit{\textbf{RQ3 (Mechanisms):} Why do agentic systems like
    \textsc{SmartOracle} work?} We investigate the advantages of 
    agentic architectures over sequential prompting and the utility 
    of semi-supervised labeling for evaluation.
\end{enumerate}

This paper makes the following 
contributions:
\begin{itemize}
    \item \textbf{Agentic Architecture:} We propose 
    \textsc{SmartOracle}, a novel agentic framework utilizing specialized 
    agents (Finder, Checker, Critic) to automate root cause analysis 
    for differential fuzzing of JavaScript.
    \item \textbf{Evaluation Methodology:} We introduce a 
    semi-supervised labeling strategy using K-means clustering to 
    efficiently establish ground truth for large-scale error logs.
    \item \textbf{Empirical Results:} We demonstrate that our 
    approach achieves 0.84 recall with an 18\% false positive rate. \textsc{SmartOracle} reduces API costs by 
    10$\times$ and is 4$\times$ faster than sequentially prompted Gemini 2.5 Pro. In active deployment, \textsc{SmartOracle} identified 
    eight previously unknown specification-level bugs.
\end{itemize}

The success of \textsc{SmartOracle}'s agentic architecture  on Javascript suggests it might be useful other software systems- a research direction we will explore in future work.

\section{Background}
\subsection{Motivation}

Fuzz testing is a cornerstone technique for identifying software vulnerabilities by executing programs against large volumes of randomly generated inputs \cite{fuzztesting, miller1990, zhu2022fuzzing}. Differential fuzzing amplifies this capability by executing multiple implementations or versions of a system on identical inputs, which surfaces bugs through discrepancies in observed behaviors \cite{mckeeman1998differential}. In practice, differential fuzzing with traditional oracles, which are predominantly based on output mismatches or simple crash checks, generates an overwhelming volume of findings, particularly as the scale of fuzzing campaigns increases \cite{bernhard2022jit}. The volume of differential findings is often proportional to the computational resources dedicated to fuzzing, thereby quickly outpacing human capacity for triage \cite{bernhard2022jit, wachter2025dumpling, lima2021}.

This high-volume problem is compounded by the large specification surface of JavaScript engines and the prevalence of undefined or implementation-dependent behaviors. Each additional core in an extensive campaign substantially increases the space of observed divergences, many of which are benign, expected, or otherwise uninteresting from a security or reliability perspective. For example, in a two-day experiment utilizing only 50 CPU cores, our experimental setup generated over 10,000 distinct differential findings when using a naive oracle, rendering exhaustive manual triage intractable.

Prior work has recognized these challenges related to scaling and noise. Bernhard et al. \cite{bernhard2022jit} emphasize the need to automate the deduplication and prioritization of findings to reduce manual effort. Wachter et al. \cite{wachter2025dumpling} stress the importance of refining oracle sensitivity to avoid spurious false positive results rooted in non-bug behavioral variation. To alleviate reliance on exhaustive human review, several studies in JavaScript engine fuzzing now employ unsupervised clustering methods to aggregate similar findings \cite{lima2021, jiang2021igor}. With clustering, a human reviewer inspects only cluster representatives, dramatically reducing the review set from thousands to a few dozen. Even so, reviewing these summaries may require non-trivial domain knowledge and remain time-consuming; in our experience, a novice reviewer may spend over 30 minutes per cluster, particularly when investigating ambiguous discrepancies.

There remains both a practical and a research imperative to reduce this human-labor bottleneck further. Automated or semi-automated oracles capable of accurately labeling true bugs and confidently filtering benign differences would fundamentally alter the cost-benefit analysis of large-scale differential fuzzing, making previously impractical campaigns feasible with limited expert attention. This motivation informs our work, in which we propose \textsc{SmartOracle}, an LLM-based agentic system designed to serve as a scalable differential oracle, reducing both the number of findings for human review and the effort required to resolve each finding.

A further challenge is evaluating new oracles. Gathering negative samples or likely false positives for robust development and benchmarking has previously required expensive manual relabeling. Cluster-then-label approaches \cite{lima2021, jiang2021igor} offer a promising compromise, enabling the creation of low-cost gold sets by focusing expert annotation only on cluster centroids. In this work, we investigate how semi-supervised label propagation methods can further accelerate the evaluation, development, and improvement of oracles for JavaScript differential fuzzing.

In summary, the scalability of differential fuzzing is currently constrained by two distinct but related human-in-the-loop bottlenecks: the semantic analysis required to filter noise and the annotation effort needed to benchmark oracles. Our work aims to dismantle both barriers. By coupling an agentic LLM-driven oracle with a semi-supervised evaluation framework, we seek to automate the suppression of benign findings and establish a rigorous, low-cost methodology for verifying such automation.

\subsection{Differential Fuzzing}

Differential fuzzing is a software testing methodology in which multiple versions or independent implementations of a system are subjected to identical, randomly-generated inputs to uncover inconsistencies in their output behavior \cite{mckeeman1998differential}. This method is particularly well-suited for systems that require conformity to a shared specification, such as programming language runtimes, compiler toolchains, or complex protocols, where behavioral equivalence is expected under all valid inputs. Typical applications include fuzzing C compilers \cite{mendoza2023grayc}, WebAssembly frameworks \cite{zhou2023wadiff}, database systems \cite{Rigger2020PQS}, cryptography libraries \cite{jung2025enhancing}, and web browsers, including JavaScript engines \cite{bernhard2022jit, wachter2025dumpling, lima2021}.

The differential fuzzing workflow comprises three stages: input generation, parallel execution on different targets, and the triage of observed discrepancies. The critical challenge resides in the final stage, which involves determining whether a discovered behavioral difference is an actual bug, an artifact of undefined behavior, or a benign difference permitted by a specification gap. Central to this step is the “differential oracle,” the logic responsible for deciding which divergences merit further investigation.

Most contemporary fuzzing setups employ naive oracles that flag any differences in standard output, program crashes, or error codes as potential bugs. While these oracles maximize sensitivity, they are prone to reporting a high number of false positives, which are differences that are either non-buggy or irrelevant, often stemming from implementation-specific choices, benign environmental effects, or allowed degrees of freedom in the specification \cite{lima2021}. To mitigate this, some works introduce static, hard-coded rules or filters into the oracle to discard common classes of known-bad findings, adjusting these filters as further real-world experience is acquired \cite{bernhard2022jit, wachter2025dumpling}.

Despite these measures, the approach remains fundamentally limited. Naive or rule-based oracles require continuous human tuning and substantial trial-and-error effort, and they still frequently overwhelm analysts with reports that require manual triage. This high human cost is the main bottleneck preventing even larger fuzzing campaigns, as the majority of findings in a typical run are ultimately not actionable bugs.

A crucial ongoing research direction thus focuses on constructing more intelligent, context-aware oracles that can automatically reason about behavioral differences, filter noise, highlight likely bugs, and scale with the increased throughput of modern fuzzing infrastructure. As execution speeds and parallelism rise, so too does the need for robust methods of deduplication, prioritization, and, ultimately, intelligent oracle feedback to guide further bug discovery.

\subsection{LLM Agents}

Large Language Model (LLM) agents are redefining the landscape of software engineering (SE) and software quality assurance (QA). In contrast to conventional approaches that treat LLMs as isolated completion engines, recent work converges on agentic models wherein LLM-powered components are delegated specific, well-defined roles, mirroring the structure of effective human engineering teams \cite{jin2024llms, hong2023metagpt}. Research in this domain demonstrates that by endowing agents with persistent memory, explicit objectives, and access to external tools or APIs, these systems can manage significantly more complex and dynamic workflows than prompt chaining alone \cite{wang2025agents, hong2023metagpt}. Consequently, LLM agents are now effectively orchestrated for end-to-end requirement gathering, code and test generation, review, debugging, and project management. State-of-the-art frameworks, such as ChatDev \cite{qian2023chatdev}, deploy agents in parallel roles including developer, reviewer, and Quality Assurance (QA) engineer to autonomously deliver working software artifacts, demonstrating both rapid output and adaptability to new project styles and requirements \cite{hong2023metagpt}.

A significant strength of the agentic paradigm is its capacity to address weaknesses endemic to purely generative LLMs, particularly concerning robustness, self-correction, and factuality \cite{castrillo2025fundamentals}. By encoding debate, critique, and review cycles among agents, these systems systematically reduce model hallucination and detect subtle errors that may be missed by less structured approaches \cite{gosmar2025hallucination}. Recent systems have demonstrated that agentic assemblies frequently outperform the sum of their parts. Frameworks employing teams of diverse LLM agents, for example, by combining tool-user, explainer, synthesizer, and critic roles, observe significant improvements in problem resolution over any individual agent baseline \cite{sreedhar2024simulating}. In practice, this multi-agent debate, task assignment, and mutual validation process facilitates the development of reliable, automated SE tooling and increases trust in automation for critical tasks such as test or patch generation and vulnerability auditing.

While initial attention was heavily focused on AI coding assistants, the scope of LLM agents in SE/QA is becoming substantially broader. Modern LLM agents are actively deployed for requirements engineering \cite{ferrari2025formal}, bug triage and root cause analysis \cite{wang2024rcagent}, coverage-oriented fuzz input generation \cite{xia2024fuzz4all}, and numerous other tasks \cite{wang2025agents}. Newer agentic architectures also exhibit flexibility in role assignment; some frameworks dynamically create or retire agents as the software development lifecycle or project complexity evolves. As the field matures, significant open questions remain regarding benchmarking, seamless human-AI collaboration, and the boundaries of “agenthood” in increasingly hybridized systems. Nevertheless, foundational research indicates a future where highly autonomous, reliable, and scalable agentic AI systems provide the backbone for continuous software quality engineering.

\subsection{Semi-supervised label propagation}

A challenge with working on Javascript is that this particular piece of
software has been extensively studied, with a comprehensive existing test
suite. Consequently, finding test cases that expose previously undetected
defects (or new defects) requires extensive exploration of increasingly subtle feature interactions and edge
cases.

In terms of machine learning, this makes JS testing a label-starved domain;
i.e., whatever we do in this domain must be done with very little data on newly discovered bugs. Label starvation complicates model building since, without good labels, it is hard to build good models.

When labels are rare, we must somehow manufacture surrogate labels from existing ones.
This is the task of {\em semi-supervised learning}. 
Semi-supervised label propagation is increasingly employed in software engineering and machine learning to expand labeled datasets efficiently by leveraging the structure and similarity of unlabeled data. Methods such as FRUGAL \cite{tu2021frugal}, co-training approaches, and code vulnerability assessment utilize feature similarity, which is often modeled using graphs where nodes represent code artifacts or bug reports and edges indicate semantic or structural proximity. Labels assigned to a small subset of high-confidence points are then propagated through these graphs, effectively bootstrapping robust classifiers using only limited manual annotation. Research by Majumder et al. \cite{majumder2024less}, among others, demonstrates that these techniques can achieve predictive performance comparable to fully supervised models, even when only 2-3\% of the dataset is labeled, through the fusion of lexical, structural, and behavioral code features.

Beyond software defect and vulnerability prediction, semi-supervised propagation is applied in bug triage, systematic literature reviews, and broader software analytics \cite{wu2017label}. Typically, a base of labeled cases, such as bugs, vulnerabilities, or citations, informs iterative algorithms that assign labels to similar unlabeled instances, often utilizing clustering, spectral embeddings, or nearest-neighbor logic. These strategies facilitate rapid scalability while controlling annotation costs, rendering them essential in domains where large volumes of raw outputs demand efficient review.

Complementary to these trends, works such as Lima et al. \cite{lima2021} have proposed clustering warnings and tool outputs as an organizational step in pipeline-based software analysis. Their approach utilizes feature-based clustering to group similar warnings, thereby facilitating human review. Unlike most semi-supervised propagation methods, however, their study did not empirically evaluate the effectiveness of the clustering algorithm itself, focusing instead on the overall system's bug-finding ability and the efficiency of downstream manual inspection. This highlights a recurring challenge: while clustering and label propagation can organize and scale annotation, quantifying their direct impact on review accuracy and triage outcomes remains an important open problem in the engineering of automated pipelines.

\subsection{Differences from previous work}


Our approach moves beyond prior work by integrating LLM-based agents for semantic triage and employing semi-supervised label propagation to scale expert annotation across discoveries. This method bridges three previously separate lines of research: advanced fuzzing sensitivity, automated semantic analysis, and scalable evaluation. Unlike previous systems that addressed these aspects in isolation, our unified framework yields broader and more automated coverage, offering meaningful gains in both efficiency and depth of bug understanding.

To understand the difference of our approach to prior work,
articles were systematically collected using highly cited paper Google Scholar 
collected from the following query:
\begin{itemize}
\item Year $\ge$ 2020; and
\begin{itemize}
\item  \texttt{differential fuzzing" AND "JavaScript"} (yielding papers 1–7 in Table ~\ref{tab:related_work}),
\item or \texttt{"software engineering" AND "LLM" AND "agent" AND "fuzz"} (papers 8–13), 
\item or \texttt{"label propagation" AND "software engineering" AND "semi supervised"} (papers 14–16). 
\end{itemize}
\end{itemize}
Any paper with single figure citations was excluded (exception: for 2025 papers, we used engineering judgment to decide if they should be included). The collected articles and their discussed subjects are summarized in Table ~\ref{tab:related_work}.

\begin{figure}
    \centering
    \includegraphics[width=0.6\linewidth]{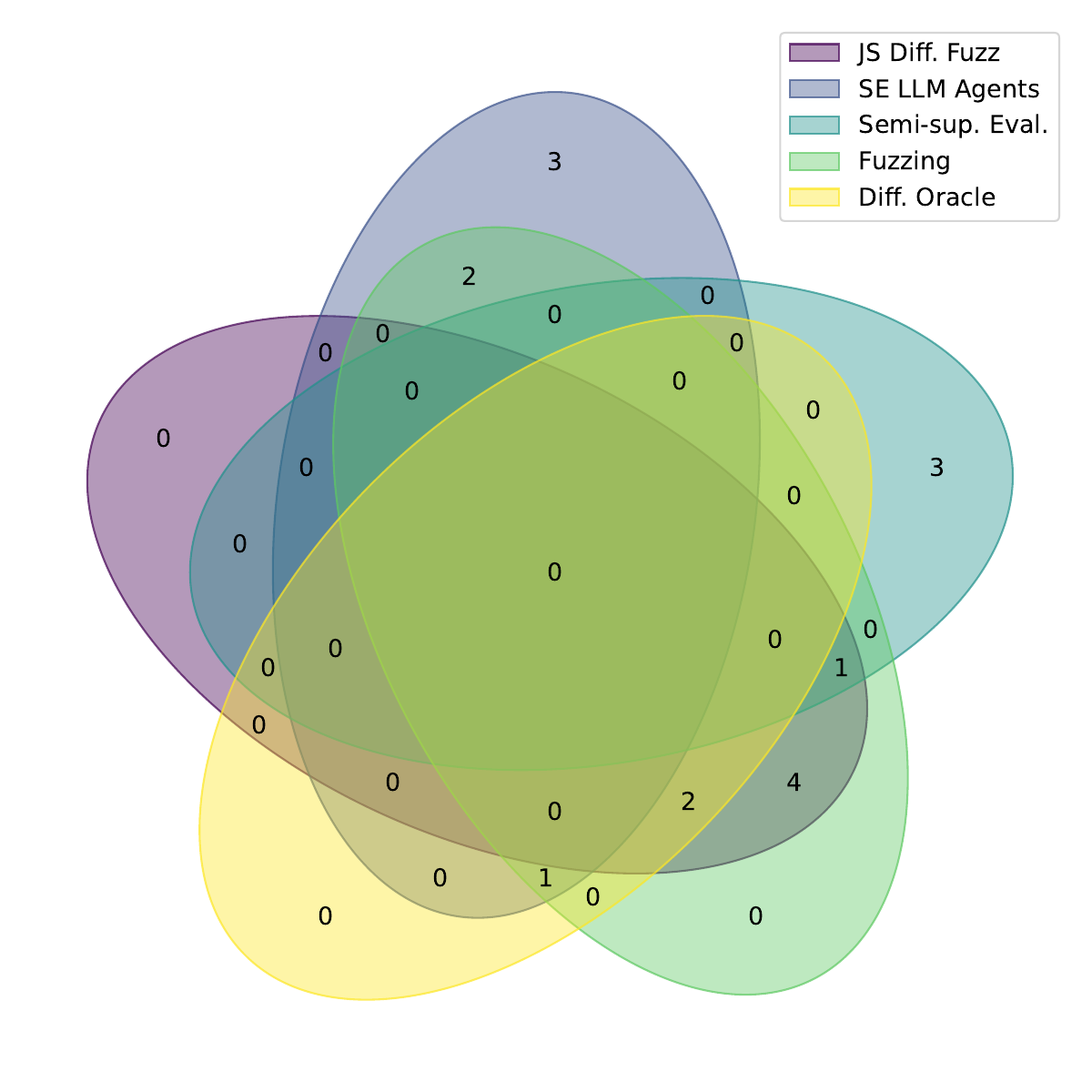}
    \caption{Intersecting related works. \textsc{SmartOracle} is the first to discuss all the related concepts}
    \label{fig:venn}
\end{figure}

As visualized in Fig. ~\ref{fig:venn}, this analysis found that no previous system has  explored the five components used in \textsc{SmartOracle}.

\begin{table}[ht]
\centering
{\small
\setlength{\tabcolsep}{2pt}
\renewcommand{\arraystretch}{1.12}
\begin{tabularx}{\textwidth}{r p{0.28\textwidth} c r c c c c c}
\hline
\textbf{\#} & \textbf{Paper} & \textbf{Year} & \textbf{Cites} & \textbf{JS Diff. Fuzz} & \textbf{SE LLM Agents} & \textbf{Semi-sup. Eval.} & \textbf{Fuzzing} & \textbf{Diff. Oracle} \\
\hline
1  & Test Transplantation (Lima et al.)\cite{lima2021} & 2021 & 17  & \checkmark & \texttimes & \checkmark & \checkmark & \texttimes \\
2  & JEST (Park et al.)\cite{jestn+1}                 & 2021 & 29  & \checkmark & \texttimes & \texttimes & \checkmark & \checkmark \\
3  & Deep Compiler Fuzzing (Ye et al.)\cite{ye2021automated}  & 2021 & 97  & \checkmark & \texttimes & \texttimes & \checkmark & \texttimes \\
4  & FuzzJIT (Ye et al.)\cite{wang2023fuzzjit}                & 2021 & 54  & \checkmark & \texttimes & \texttimes & \checkmark & \texttimes \\
5  & JIT Picking (Bernhard et al.)\cite{jitpicking}      & 2022 & 76  & \checkmark & \texttimes & \texttimes & \checkmark & \texttimes \\
6  & AccuOracle (Li et al.)\cite{accuoracle}             & 2025 & 0   & \checkmark & \texttimes & \texttimes & \checkmark & \checkmark \\
7  & DUMPLING (W\"{a}chter et al.)\cite{wachter2025dumpling}          & 2025 & 5   & \checkmark & \texttimes & \texttimes & \checkmark & \texttimes \\
\hline
8  & Hu et al.\cite{hu2023large}                          & 2023 & 97  & \texttimes & \checkmark & \texttimes & \texttimes & \texttimes \\
9  & Xia et al.\cite{xia2024fuzz4all}                         & 2024 & 283 & \texttimes & \checkmark & \texttimes & \checkmark & \texttimes \\
10 & Mao et al.\cite{mao2024multi}                         & 2024 & 20  & \texttimes & \checkmark & \texttimes & \texttimes & \texttimes \\
11 & Wang et al.\cite{wang2024xuat}                        & 2024 & 82  & \texttimes & \checkmark & \texttimes & \texttimes & \texttimes \\
12 & Amusuo et al.\cite{amusuo2025falsecrashreducer}                      & 2025 & 0   & \texttimes & \checkmark & \texttimes & \checkmark & \texttimes \\
13 & Xu et al.\cite{xu2025ckgfuzzer}                          & 2025 & 8   & \texttimes & \checkmark & \texttimes & \checkmark & \checkmark \\
\hline
14 & Tu et al.\cite{tu2021frugal}                          & 2021 & 29  & \texttimes & \texttimes & \checkmark & \texttimes & \texttimes \\
15 & Majumder et al.\cite{majumder2024less}                    & 2024 & 11  & \texttimes & \texttimes & \checkmark & \texttimes & \texttimes \\
16 & Pei et al.\cite{pei2025semi}                         & 2025 & 0   & \texttimes & \texttimes & \checkmark & \texttimes & \texttimes \\
\hline
   & \textbf{\textsc{SmartOracle} (Our work)}   & 2025 &  -   & \checkmark & \checkmark & \checkmark & \checkmark & \checkmark \\
\hline
\end{tabularx}
}
\caption{Comparison of related work (abbreviations: JS Diff. Fuzz = JavaScript Differential Fuzzing, LLM Agents = Software Engineering LLM Agents, Semi-sup. Eval = Semi-supervised Evaluation, Diff. Oracle = Differential Oracle).}
\label{tab:related_work}
\end{table}

\section{Experimental Methodology}

Our methodology commences with the construction of benchmark datasets, which include historical differential JavaScript engine bugs and manually labeled new findings generated using mutational fuzzing campaigns across four major engines. \textsc{SmartOracle} is then systematically evaluated for its ability to detect true bugs, filter false positives, and generalize across real-world test cases. We present direct comparisons to sequential LLM prompting workflows and conduct further analysis of triage accuracy, runtime efficiency, and cost effectiveness. Finally, to minimize human labeling effort, we incorporate a semi-supervised label propagation technique that facilitates targeted manual review to efficiently expand ground truth sets. 

\subsection{Datasets}

\subsubsection{Historical Differential Fuzzing Data}

\begin{figure}
    \centering
    \includegraphics[width=\linewidth]{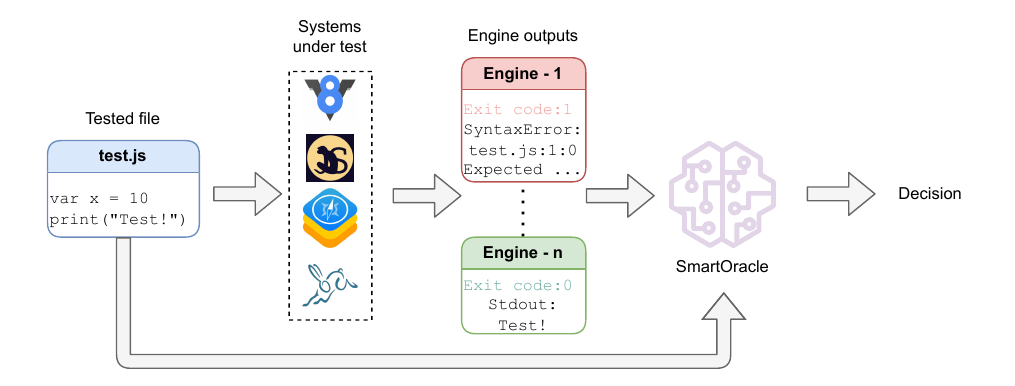}
    \caption{Overview of how a single test file is processed and the differential findings are analyzed by \textsc{SmartOracle}}
    \label{fig:smartoracle_flow}
\end{figure}

To test \textsc{SmartOracle}, the workflow described in Fig. \ref{fig:smartoracle_flow} is followed. First, logs are obtained from each of the engines being differentially tested. These logs are passed to \textsc{SmartOracle}, along with the code that exhibits this behavior across these engines. SmartOracle provides a decision on whether this bug is reportable or needs to be skipped.

The experiments using historical data were designed to assess SmartOracle’s performance using a standardized set of differential JavaScript engine bugs documented in prior research. We utilize bugs reported in two prior JS differential fuzzing papers: Park et al. \cite{jestn+1} and Lima et al. \cite{lima2021}. To faithfully reproduce the conditions of prior studies, legacy versions of all engines involved were installed and configured in the test environment. SmartOracle’s evaluation workflows and its terminal runner tool were modified to match the runtime setups described in the original work precisely. 

The Park et al. dataset comprises 44 confirmed bug cases originally discovered through differential fuzzing across four major engines: V8 \cite{v8js}, GraalJS \cite{graaljs}, QJS \cite{quickjs}, and ModdableXS \cite{moddablexs}. The code snippets were executed using the test harness provided in their GitHub repository. The Lima et al. dataset comprises 31 confirmed bug cases discovered through differential fuzzing across four major engines: V8, SpiderMonkey \cite{spidermonkey}, JavsScriptCore \cite{javascriptcore}, and  ChakraCore \cite{chakracore}. We were able to reproduce behavior on 16 of the 31 bugs. We were unable to procure usable differential findings for the rest due to unavailable test harnesses for the tested engines.

The workflow to reproduce these bugs were:

\begin{itemize}
\item The original code sample was executed in all targeted engines to generate fresh output logs.
\item \textsc{SmartOracle} received these logs along with the code, simulating a real-world engine triage scenario.
\item The system’s orchestrator coordinated inputs to sub-agents, which scrutinized behavioral discrepancies, checked compliance against the ECMAScript specification \cite{ECMA262_2025}, and verified against patterns of known false positives and duplicates. 
\end{itemize}

The outcome from \textsc{SmartOracle} is a verdict to either REPORT (if a specification violation was believed to be present) or SKIP. To quantify accuracy, recall was calculated as the fraction of ground-truth bugs correctly reported by \textsc{SmartOracle} out of the total number of bugs.

\subsubsection{Manually Labeled Fuzzing Data}
\label{sec:manually_labeled}

Using findings from our own fuzzing sessions to test \textsc{SmartOracle} or the semi-supervised labeling would not be possible without proper labeling. To address this, the lead author manually labeled 710 findings from a short differential fuzzing session. The findings were labeled as either BUG or NO\_BUG, along with a one-word root cause.

The dataset finally comprised 710 findings, out of which 518 were bugs and 192 did not contain bugs. It is worth noting that this imbalance is not representative of all differential testing campaigns. The seeds of this campaign were bug-invoking JS test files, which led to several similar buggy findings. The primary purpose of this dataset was to evaluate \textsc{SmartOracle}, which required adequate buggy findings to be tested.

\subsection{Sequential LLM Prompting}

A significant open question in differential testing workflows is whether agentic approaches, comprised of orchestrated sub-agents and dynamic tool use, offer concrete advantages over standard sequential LLM prompting, which relies on stepwise, Chain-of-Thought execution within a single Large Reasoning Model (LRM). To address this, an experiment was designed to compare \textsc{SmartOracle} against a sequentially prompted LRM on the benchmark differential bug dataset.

\begin{figure}
    \centering
    \includegraphics[width=0.5\linewidth]{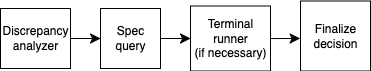}
    \caption{Sequentially prompted LRM setup}
    \label{fig:sequential_llm}
\end{figure}

In the sequential setup, as illustrated in Fig. \ref{fig:sequential_llm}, a single LRM is prompted in a fixed sequence intended to mirror the core reasoning steps performed by \textsc{SmartOracle}'s sub-agents. The sequence proceeds as follows: (1) the LRM analyzes and summarizes engine discrepancies; (2) it queries relevant sections of the ECMAScript specification for compliance; (3) it determines if execution in the engine terminal is necessary for further verification; and (4) it synthesizes these results to reach a final bug-finding decision. This pipeline intentionally excludes duplicate detection to focus the comparison on the primary bug identification capabilities of both methodologies.

The objective of this design is to emulate the decision quality achievable by specialized sub-agents through conventional prompt chaining, while minimizing additional complexity. Consistent with contemporary findings, this method demonstrates how sequential LLM workflows maintain explicit control and predictability, yet lack the dynamic orchestration and autonomous tool selection characteristic of modern agentic systems. Notably, recent work by Belcak et al. \cite{belcak2025small} argues for the efficiency of small agentic systems over monolithic, sequentially prompted LRMs, thereby supporting the relevance of this comparison in evaluating real-world tradeoffs.
Both \textsc{SmartOracle} and the sequential LRM were evaluated across the same dataset, allowing for a direct measurement of recall, resource utilization, and overall triage effectiveness under two distinct reasoning paradigms.

\subsubsection{Rationale for model architecture selection}
\begin{wrapfigure}{r}{2in}
    \centering
\includegraphics[width=2in]{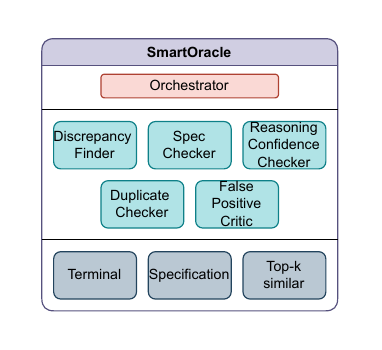}
    \caption{Overview of SmartOracle, its sub-agents, and tools.}
    \label{fig:agent}
\end{wrapfigure} 

For the sequential LLM baseline, a single large reasoning model (LRM), Gemini 2.5 Pro, was selected to maximize generality and reasoning depth within a single stepwise workflow. By contrast, the agentic setup of SmartOracle utilizes multiple instances of Gemini 2.5 Flash as task-specialized sub-agents.

We draw upon the findings of Belcak et al., who demonstrate that smaller, specialized models often suffice for scoped agentic tasks while offering superior speed and economy. While the exact parameter counts of closed-source commercial models, such as Gemini 2.5 Flash, are proprietary and may exceed the strict definitions of "Small" Language Models (SLMs) \cite{IBM_SLM_2025}, they are explicitly engineered as high-efficiency, low-latency alternatives to their "Pro" counterparts. In this study, Flash occupies the functional role of an SLM. It provides the requisite throughput for multi-turn agentic loops, while the Pro model serves as the heavyweight reasoning baseline.

\subsection{Agent Design}

Our agentic system comprises a central orchestrator, five specialized sub-agents, and three integrated tool interfaces. The orchestrator coordinates the workflow, dynamically delegating tasks to sub-agents based on the complexity and type of issue encountered, in accordance with recent advancements in LLM agent research. Figure \ref{fig:agent} depicts the overall agent structure. It is worth noting that each sub-agent is permitted to utilize a subset of the tools and is also allowed to invoke other relevant sub-agents.

The prompts for each sub-agent are designed using reasoning pathways distilled from extensive sequential Chain-of-Thought (CoT) prompting with Large Reasoning Models (LRMs) \cite{jaech2024openai}, reflecting patterns commonly observed in decision-making logs generated by LRMs acting as differential oracles. For brevity, only representative prompt examples for the orchestrator and one sub-agent are shown.

The prompts used by the sub-agents, along with the purpose of each agent and tool, are detailed below. For brevity, only the prompts for the orchestrator and a single sub-agent are presented.

\subsubsection{Orchestrator}

The orchestrator’s primary role is to parse and structure the task, then execute bug identification by consulting available sub-agents and the relevant tool interfaces. It is responsible for generating the final verdict for each finding, categorized as either REPORT (actual bug) or SKIP (not a true bug). The specific instructions provided to the orchestrator are illustrated in Figure \ref{prompt:orch}.

\begin{figure}[ht]
\begin{tcolorbox}[title=Prompt: Orchestrator]
You are a JavaScript engine differential testing triage agent.

Analyze the provided finding and determine whether to REPORT or SKIP, including your decision with confidence and rationale.

\textbf{Workflow:}
\begin{enumerate}
  \item Summarize the issue briefly and neutrally.
  \item Identify and understand the behavioral discrepancy in the data.
  \item Use sub-agents as needed:
  \begin{itemize}
    \item Discrepancy structurer: Structure the behavioral difference.
    \item Specification checker: Evaluate ECMA-262 compliance.
    \item Test case minimizer: Create minimal reproducible examples.
    \item Duplicate analyzer: Check for known issues.
    \item False positive critic: Identify false positives.
  \end{itemize}
  \item Use tools (such as terminal, specification, and engine queries) if appropriate.
  \item Synthesize findings from sub-agents for a comprehensive assessment.
  \item Make a decision with confidence and rationale.
\end{enumerate}

\textbf{Decision criteria:}
\begin{itemize}
  \item \textbf{REPORT:} Clearly violates ECMA-262 specification.
  \item \textbf{SKIP:} False positive, duplicate, or not a specification violation.
  \item \textbf{Confidence:} Scale from 0.0 to 1.0, based on evidence.
  \item \textbf{Rationale:} Concise explanation.
\end{itemize}

At the end, ensure your results use the final decision with confidence and rationale. Consult the false positive critic for a maintained list of typical non-reportable patterns. Be concise and focused on the core issue.
\end{tcolorbox}
\caption{Prompt to the orchestrator LLM}
\label{prompt:orch}
\end{figure}

\subsubsection{Discrepancy Finder}

The discrepancy finder sub-agent is tasked with recognizing and formalizing behavioral divergences observed between different JavaScript engine outputs. It receives a system-level instruction to thoroughly analyze the discrepancy, summarize its key characteristics, and propose plausible root causes. Access to tools such as specification queries and in-context code execution allows the sub-agent to enhance decision quality. The detailed prompt for the discrepancy analyzer is given in Figure \ref{fig:discrepancy-analyzer-prompt}.

\begin{figure}[ht]
\centering
\begin{tcolorbox}[title=Prompt: Discrepancy Finder Sub-Agent]
Extract, normalize, and structure the behavioral discrepancy across engines.

Summarize the divergence and propose likely root cause candidates.

Provide a minimal, self-contained code snippet that reliably reproduces the discrepancy.

\textbf{Available tools:}
\begin{itemize}
    \item terminal(engine\_name, code): Run JavaScript code on configured engines.
    \item spec(query): Search ECMA-262 specification for relevant sections.
\end{itemize}

Use tools as needed to better understand the discrepancy.
\end{tcolorbox}
\caption{Prompt to the discrepancy analyzer sub-agent LLM}
\label{fig:discrepancy-analyzer-prompt}
\end{figure}

\subsubsection{Specification Checker}

The specification sub-agent serves to ground the reasoning process in the ground truth of the JavaScript language specification. The language specification is cached locally and exposed to the sub-agent as a querying tool. The sub-agent is prompted to query the specification based on its analysis of the identified discrepancy or issue. This sub-agent provides the relevant portion of the specification pertaining to the issue in a readable format.

\subsubsection{Reasoning Confidence Checker}

At the conclusion of the analysis, a confidence score is assigned to the reasoning used to arrive at a decision. The reasoning confidence checker is prompted to evaluate the evidence and provide a quantitative measure of confidence for the decision rendered by the agent. This metric is important for determining whether a decision should be retained or discarded.

\subsubsection{Duplicate Checker}

As the agent processes the corpus of differential findings, avoiding the report of duplicate bugs becomes a significant problem. Each time the agent reports a bug, the decision is stored alongside a concise summary. The duplicate checker sub-agent is prompted to utilize the top-k similar issue tool to verify if the current issue has been previously observed. If the issue is identified as a duplicate, the duplicate checker instructs the orchestrator to SKIP the finding.

\subsubsection{False Positive Critic}

The false positive critic sub-agent is prompted to critique the reasoning process, aiming to avoid common false positive issues and logical pitfalls. This sub-agent ensures that the reasoning process is grounded in a quantifiable specification deviation rather than a superficial output difference. JavaScript is a language that contains several parts of its specification with undefined behavior, allowing engine developers to implement their own solutions. Such undefined behavior cannot be reported as a genuine issue and must be treated as a false positive.

\subsubsection{Tools}

Three tools are exposed to each of the agents:
\begin{itemize}
    \item \textit{The terminal tool} allows the agents to run a JavaScript code snippet against any of the tested engines. The engines are pre-installed and exposed to the agent through an MCP server \cite{ModelContextProtocol2025}
    \item \textit{The specification tool} allows sub-agents to query the specification to ground the reasoning based on truth from the defined behavior in the specification.
    \item \textit{The top-k similar tool} provides a list of \(k\) similar (\(k=10\) by default) issues that the agent has reported to avoid re-reporting issues.
\end{itemize}

\subsection{Current Version bug finding}

Differential fuzzing experiments are conducted on four major JavaScript engines: V8, SpiderMonkey \cite{spidermonkey}, JavaScriptCore \cite{javascriptcore}, and GraalJS, as detailed in Table \ref{tab:js-engines}. These engines are collectively referred to as the Systems Under Test (SUT). The fuzzer utilized for this study is KITTEN \cite{xie2025kitten}, which performs mutational fuzzing using seeds obtained from the DIE corpus \cite{park2020fuzzing}.

KITTEN is selected due to its novel design, which effectively discovers bugs in JavaScript engines. The DIE corpus provides a rigorously curated repository of JavaScript code snippets, including known vulnerabilities and regression tests drawn from major engine projects. This combination ensures broad and meaningful coverage of potential behavioral discrepancies.

The fuzzing campaign was executed for 48 continuous hours, applying mutational fuzzing from the DIE seed set against all four SUTs. This process yielded approximately 10,000 differential findings, defined as cases where the STDOUT or STDERR output of at least one engine differs from that of the others. Differential findings where all engines fail identically or produce the same output are excluded from consideration.

To manage the large volume of findings, hierarchical clustering is applied. This process involves initially grouping findings by exit code, followed by a finer-grained clustering based on the TF-IDF \cite{tf-idf} encoding of the textual content within each finding's output logs. This multi-level clustering strategy reduces the manual review burden by focusing attention on distinct and representative behavioral differences.

\begin{table}[ht]
\centering
\begin{tabular}{lll}
\hline
Engine           & Developer/Project      & Used In (Main Products)      \\
\hline
V8               & Google/Chromium Project  & Chrome\cite{chromium08}, Node.js\cite{nodejs09}, Electron\cite{electron13}    \\
JavaScriptCore   & Apple/WebKit    & Safari\cite{safari03}, WebKit Apps          \\
SpiderMonkey     & Mozilla Foundation     & Firefox\cite{firefox02}, Servo\cite{servo12}   \\
GraalJS          & Oracle/GraalVM      & GraalVM\cite{graalvm18}, Java apps, DB engines \\
\hline
\end{tabular}
\caption{Basic information about four JavaScript engines used for differential fuzzing experiments.}
\label{tab:js-engines}
\end{table}

\subsubsection{Coverage}

Coverage metrics were not collected or reported in our study. Although coverage is commonly used for guiding and evaluating fuzzing, it is only weakly correlated with effective bug discovery in differential fuzzing contexts ~\cite{klees2018evaluating, hur2021difuzzrtl}. As \textsc{SmartOracle} is complementary and agnostic to the underlying fuzzing tool, it can be applied in any differential testing setup regardless of the coverage achieved. Collecting detailed coverage data would add complexity unrelated to our primary focus on bug identification and semantic triage.

\subsection{Semi-supervised Evaluation Methodology}
\label{sec:semisupervised_method}

To determine if clustering can effectively separate differential findings for efficient triage, we utilized the manually labeled dataset of 710 findings described in Section ~\ref{sec:manually_labeled}. We evaluated our approach at two levels of granularity: (1) Binary Classification, distinguishing between buggy and benign findings; and (2) Root Cause Categorization, identifying specific failure modes.

\subsubsection{Feature Extraction}
We transformed the raw textual logs into numerical feature vectors using the Term Frequency-Inverse Document Frequency (TF-IDF) method. This technique weighs terms by their distinctiveness, filtering out common stop words while emphasizing terms unique to specific error types. For a term $t$, document $d$, and corpus $D$, the weight is calculated as:

\begin{equation}
    W_{t,d} = \text{tf}(t,d) \times \log \left( \frac{N}{|\{d \in D : t \in d\}|} \right)
\end{equation}

where:
\begin{itemize}
    \item $\text{tf}(t,d)$ is the frequency of term $t$ in document $d$.
    \item $N$ is the total number of documents in the corpus.
    \item $|\{d \in D : t \in d\}|$ is the number of documents containing term $t$.
\end{itemize}

\subsubsection{Clustering Strategies and Optimization}
We compared two grouping strategies:
\begin{enumerate}
    \item \textbf{Exit Code Grouping:} Findings are grouped solely by their exit code tuple (the exit status across all four engines).
    \item \textbf{K-means Refinement:} We applied K-means clustering within each exit code group. To determine the optimal number of clusters $k$ without using ground-truth labels, we utilized the Elbow Method based on the Silhouette Coefficient. 
\end{enumerate}

The Silhouette Coefficient $s(i)$ for a single data point $i$ measures how similar it is to its own cluster compared to other clusters:

\begin{equation}
    s(i) = \frac{b(i) - a(i)}{\max\{a(i), b(i)\}}
\end{equation}

where:
\begin{itemize}
    \item $a(i)$ is the mean distance between $i$ and all other points in the same cluster (intra-cluster distance).
    \item $b(i)$ is the mean distance between $i$ and all points in the nearest neighboring cluster (inter-cluster distance).
\end{itemize}

We selected the $k$ that maximized the average silhouette score across all points in the pattern.

\subsubsection{Medoid-Based Label Propagation}
We employed a medoid-based propagation strategy. For a cluster $C$ containing $n$ findings with feature vectors $\{\mathbf{v}_1, \ldots, \mathbf{v}_n\}$, we compute the centroid $\mathbf{c}$ as the mean vector. The medoid $\mathbf{m}$ is identified as the finding minimizing the cosine distance to the centroid:

\begin{equation}
    \mathbf{m} = \arg\min_{\mathbf{v} \in C} \left( 1 - \frac{\mathbf{c} \cdot \mathbf{v}}{||\mathbf{c}|| \cdot ||\mathbf{v}||} \right)
\end{equation}

The ground-truth label of this medoid is propagated to all other findings in the cluster.

\subsubsection{Evaluation Metric}
We measure the effectiveness of the semi-supervised approach using Propagation Accuracy, defined as the percentage of findings where the propagated medoid label matches the manual ground truth label.

\section{Results} 

The results derived from the experimental methods described in the preceding section are presented below, corresponding to the relevant research questions.

\subsection{RQ1(a): Can \textsc{SmartOracle} uncover true differential testing bugs?}\label{table3} To rigorously evaluate \textsc{SmartOracle}’s bug-finding ability, we employ two datasets of previously reported bugs, documented in prior differential fuzzing investigations. Additionally, we test \textsc{SmartOracle} on our manually labeled dataset from a brief fuzzing campaign. It is worth noting that our manually labeled dataset may contain duplicates, i.e., the same bug could manifest through different JavaScript input files. This redundancy is addressed in practice using the duplicate analyzer sub-agent; however, to evaluate \textsc{SmartOracle}'s bug-finding ability, we disable the duplicate analyzer sub-agent and allow it to process all findings as-is.

The principal evaluation metric is recall, as shown in Equation ~\ref{eqn:recall}, which, given an entirely positive dataset, measures the proportion of actual bugs correctly identified as such: \begin{equation} Recall=\frac{True Positives}{TruePositives+FalseNegatives} \label{eqn:recall} \end{equation} \textsc{SmartOracle} reported 84\% recall (0.84), correctly identifying 37 out of 44 bugs in the Park et al dataset. The performance is slightly lower on the Lima et al. dataset and the manually labeled data, with recalls in the 0.7 range. We attribute the higher recall in the Park et al. dataset to the assertions in the input files, which provide concrete evidence for discrepancies. \textsc{SmartOracle} may use the failed assertions to support its reasoning for bugs in the Park et al. dataset.

\begin{table}[h]
\centering
\begin{tabular}{lcc}
\hline
Dataset & Number of bugs & Recall \\
\hline
Park et al. & 44 & 0.84 \\
Lima et al. & 16 & 0.75 \\
Manually labeled & 238 & 0.73 \\
\hline
\end{tabular}
\caption{Bug finding ability of \textsc{SmartOracle} on historical and manually labeled datasets}
\label{tab:bug_finding}
\end{table}

\begin{center}
\begin{tcolorbox}[colback=gray!7, colframe=gray!80!black, title=Answer to RQ1(a):]
\textsc{SmartOracle} demonstrates excellent performance in identifying bugs manifested through differential fuzzing.
\end{tcolorbox}
\end{center}

\subsection{RQ1(b): Can \textsc{SmartOracle} understand when a finding may not be a reportable bug?} Addressing non-reportable findings is a critical component of reliable differential fuzzing. To evaluate \textsc{SmartOracle}’s ability to avoid false positives, defined as the suppression of benign divergences to flag only actionable bugs, the standard false positive rate metric is adopted: \begin{equation} FalsePositiveRate=\frac{FalsePositives}{FalsePositives+TrueNegatives} \label{eqn:fpr} \end{equation}

To establish a ground truth of negative fuzz findings, we use the manually labeled dataset detailed in Section \ref{sec:manually_labeled}. We use 136 findings as negatives or differential behavior that is non-reportable. We tested \textsc{SmartOracle} on these 136 examples and found a False Positive rate of 18\%. That is, \textsc{SmartOracle} reported 24 out of the 136 negative findings.

In summary, \textsc{SmartOracle} demonstrates significant progress toward minimizing false positives in differential fuzzing. The system achieves a low error rate while reducing the manual burden, an essential contribution for any practical triage pipeline in high-volume fuzz campaigns.

\begin{center} \begin{tcolorbox}[colback=gray!7, colframe=gray!80!black, title=Answer to RQ1(b):] \textsc{SmartOracle} can suppress the majority of non-actionable findings, achieving a low false positive rate while streamlining human review. \end{tcolorbox} \end{center}

\subsection{RQ2: Can agentic systems like \textsc{SmartOracle} 
    discover novel bugs in the latest versions of major JavaScript 
    runtimes?} To assess SmartOracle's ability to identify new, actionable bugs in current JavaScript engines, it was applied to the latest versions of GraalJS, V8, and JavaScriptCore. Across these experiments, SmartOracle identified eight distinct, previously unreported differential bugs spanning parser logic, type enforcement, and global object behaviors. Table ~\ref{tab:js-engine-bug-ids-resolution} summarizes the reproducible bug reports and their subsequent dispositions.

Of the eight candidate bugs, four were confirmed by the respective runtime developers and judged to be genuine specification violations or undesirable deviations, demonstrating that \textsc{SmartOracle}’s findings are both novel and relevant for engineering review. In one case, a parsing bug (GraalJS \#931) resulted in an upstream fix, thereby underlining its real-world impact. Three additional V8 findings were marked as low priority but acknowledged as either nonconformant behavior or non-uniform behavior across the shell and web versions of the engine.

For four JavaScriptCore bugs, developer confirmation was pending at the time of writing, illustrating the real-world challenge of timely, community-driven triage and validation. Notably, the spectrum of issues reported, ranging from unintended TypeErrors to shells with undocumented file system side effects, highlights \textsc{SmartOracle}'s capacity for identifying deep behavioral inconsistencies that baseline oracles may not capture.

This pattern of rapid bug discovery across evolving and complex JavaScript engines validates the practical value of an agentic, LLM-driven triage system for continuous and differential fuzz testing in contemporary language runtimes. By automating semantic log analysis and providing clear, actionable summaries, \textsc{SmartOracle} bridges the gap between high-volume input generation and the nuanced bug reporting required for real-world impact.

\begin{table}[ht]
\centering
\begin{tabular}{rlp{6cm}lll}
\hline
\textbf{\#} & \textbf{Engine} & \textbf{Bug Summary} & \textbf{Issue ID} & \textbf{Status} & \textbf{Resolution} \\
\hline
1 & GraalJS & Parsing Bug: Incorrect SyntaxError due to ASI within an object literal following a return statement & 931 & confirmed & fixed \\
\hline
2 & V8 & Assignment to Object.prototype.\_\_proto\_\_ does not throw a TypeError & 438787152 & confirmed & low priority \\
3 & V8 & d8: Applying 'new' to a print() call does not throw a catchable TypeError & 446261067 & confirmed & low priority \\
4 & V8 & d8 shell: Non-standard write function lacks input validation, leading to masked errors & 446661133 & confirmed & low priority \\
\hline
5 & JSC & String.prototype.search does not throw TypeError for non-callable [Symbol.search] & 297423 & unconfirmed & - \\
6 & JSC & JSC throws a TypeError when evaluating a standalone \_\_proto\_\_ identifier in the global scope & 298112 & unconfirmed & - \\
7 & JSC & jsc shell's global write function has undocumented file system side effects and masks standard errors & 299296 & unconfirmed & - \\
8 & JSC & \_\_proto\_\_ access on the global object throws TypeError & 299441 & unconfirmed & - \\
\hline
\end{tabular}
\caption{Summary of reported bugs by engine}
\label{tab:js-engine-bug-ids-resolution}
\end{table}

\begin{center}
\begin{tcolorbox}[colback=gray!7, colframe=gray!80!black, title=Answer to RQ2:]
\textsc{SmartOracle} reliably discovers and reports genuine bugs in up-to-date JavaScript engines, demonstrating effectiveness beyond historical datasets and direct utility for runtime maintainers.
\end{tcolorbox}
\end{center}

\subsection{RQ3(a): Is there an advantage in using \textsc{SmartOracle} over a sequentially prompted LRM for this task?} To quantify the comparative advantages of agentic versus monolithic LLM workflows in triaging differential fuzzing findings, parallel evaluations were performed using SmartOracle (which utilizes multi-role sub-agents and integrated tool support) and a baseline workflow employing a single LRM with sequential Chain-of-Thought (CoT) prompting. Both systems processed an identical benchmark set of historical JavaScript engine bugs to ensure a fair comparison. We deliberately chose this comparison to highlight two distinct architectural benefits.

First, this comparison tests the hypothesis that agentic orchestration can bridge the reasoning gap of smaller models. By demonstrating that the smaller-model based SmartOracle achieves higher recall than the significantly larger baseline LRM, we validate that the performance gains are driven by the framework's specialized sub-routines rather than raw model parameter counts.

Second, the reported reductions in token usage and cost reflect the architectural advantage of conditional execution. Unlike the sequential approach, which must linearly process the full reasoning context for every finding, the agentic architecture dynamically selects tools and sub-agents as complexity changes. To ensure that token costs are compared fairly, our prompts were designed such that the sum of the lengths of all sub-agent prompts is longer than the prompts in the sequential approach. Reduced token usage, therefore, represents a clear win for the agentic approach, driven by efficient execution paths rather than prompt brevity.

\begin{table}[ht]
\centering
\begin{tabular}{lcccc}
\hline
\textbf{Method} & \textbf{Recall} & \textbf{Time/Case(s)} & \textbf{Tokens/Case(/1000)\footnotemark[1]} & \textbf{Cost/Case(in USD)\footnotemark[2]} \\
\hline
\textsc{SmartOracle} &0.84  &20  &6.5  & 0.003 \\
LRM + CoT   &0.68  &91  &14.7  & 0.04 \\
\hline
\end{tabular}
\caption{Comparison between \textsc{SmartOracle} and LRM + CoT on recall, execution time, token usage, and cost per case.}
\label{tab:oracle_comparison}
\end{table}

\footnotetext[1]{Total tokens. Cost of input and output tokens may differ.}
\footnotetext[2]{As of 28th October 2025 \url{https://ai.google.dev/gemini-api/docs/pricing}. Cost calculated is total input + output token cost.}

As shown in Table ~\ref{tab:oracle_comparison}, the agentic \textsc{SmartOracle} system exhibits higher recall, indicating that its distributed, tool-augmented workflow more reliably identifies subtle specification violations and distinguishes true bugs from noise.

Significantly, the performance gap extends beyond accuracy. \textsc{SmartOracle} achieves a 4x reduction in mean analysis time and a 10x decrease in cost per finding. Specifically, the multi-agent approach averaged 20 seconds and 6.5 tokens per case (\$0.003), compared to 91 seconds and 14.7 tokens (\$0.004) for the sequential LRM. This substantial resource efficiency translates directly into improved scalability for industrial or continuous fuzzing campaigns, enabling practitioners to triage larger corpora of differential outputs without a proportional increase in cost or run time.

Beyond these empirical metrics, the structured agentic approach offers qualitative benefits. Its modular decomposition facilitates the systematic enhancement and debugging of individual analysis steps (e.g., discrepancy normalization, spec-checking, duplicate suppression), allowing the system to adapt more readily to new engine versions or triage targets.

\begin{center} \begin{tcolorbox}[colback=gray!7, colframe=gray!80!black, title=Answer to RQ3(a):] \textsc{SmartOracle} outperforms the sequentially prompted LRM across all evaluated metrics as a differential oracle. Its agentic architecture enables smaller models to achieve higher recall through specialized orchestration, while dynamic execution paths significantly reduce computational cost and time. 
\end{tcolorbox} \end{center}

\subsection{RQ3(b): Can semi-supervised label assignment aid in the evaluation of differential oracles?}
\label{sec:rq3b_results}

The method described above relies on semi-supervised label propagation.
This section describes a small secondary study that investigates the
value of label propagation over fuzzing data. In this study,
we applied the methodology described in Section ~\ref{sec:semisupervised_method} to evaluate the accuracy of semi-supervised labeling for both binary bug detection and granular root cause analysis.

\subsubsection{Binary Classification Results}
We first examined the effectiveness of grouping findings solely by their exit code patterns for detecting bugs. Out of 710 labeled findings, 98.7\% were correctly classified as BUG or NO\_BUG simply by propagating the label of the medoid from each exit code pattern. This indicates that exit code patterns act as a highly effective first-level filter, as genuine specification violations typically produce execution states distinct from benign differences.

\subsubsection{Root Cause Analysis Results}
While exit codes suffice for binary detection, practical triage requires distinguishing between specific root causes. Our dataset of 710 findings spans 14 unique exit code patterns and contains 31 distinct root cause categories. Findings with identical exit codes often stem from distinct underlying issues; for instance, a generic crash exit code might mask multiple unique failure modes. 

We compared the accuracy of propagating granular root cause labels using the baseline exit code grouping versus the K-means refinement. The baseline strategy, which treats each of the 14 exit code patterns as a single cluster, achieved an average propagation accuracy of 77.34\%. By applying K-means clustering with unsupervised $k$-selection, the accuracy improved to 89.15\%, an increase of 11.81 percentage points.

The performance gains were most pronounced in patterns exhibiting high root cause diversity. Table ~\ref{tab:pattern_analysis} details the performance on specific patterns. For example, pattern\_7\_3\_0\_1 contained 9 distinct root causes across 14 findings. The baseline approach, which assigns a single label to the group, achieved an accuracy of only 28.57\%. K-means successfully identified the internal structure, selecting an optimal $k=8$ and improving accuracy to 92.86\%.

\begin{table}[h]
\centering
\caption{Root cause propagation accuracy on select representative patterns}
\label{tab:pattern_analysis}
\small
\begin{tabular}{lcccc}
\toprule
Pattern & Baseline & K-means & Opt. $k$ & \% Gain \\
\midrule
pattern\_7\_3\_0\_1 & 28.57\% & 92.86\% & 8 & +64.29 \\
pattern\_0\_3\_3\_1 & 50.00\% & 75.00\% & 2 & +25.00 \\
pattern\_7\_3\_3\_0 & 53.85\% & 76.92\% & 4 & +23.08 \\
pattern\_0\_0\_0\_0 & 60.00\% & 80.00\% & 2 & +20.00 \\
pattern\_7\_0\_3\_0 & 98.90\% & 99.34\% & 16 & +0.44 \\
pattern\_7\_0\_0\_0 & 96.77\% & 96.61\% & 20 & -0.16 \\
\bottomrule
\end{tabular}
\end{table}

Conversely, for patterns that were already dominated by a single root cause, such as pattern\_7\_0\_0\_0, K-means maintained the high baseline accuracy within a negligible margin. This demonstrates that the unsupervised Elbow Method effectively avoids over-clustering when the data is already homogeneous.

These results highlight that while exit code patterns are necessary for high-level triage, they are insufficient for root cause analysis. Textual clustering using TF-IDF provides the semantic resolution necessary to distinguish between distinct failure modes that result in identical engine exit states.

\begin{center} \begin{tcolorbox}[colback=gray!7, colframe=gray!80!black, title=Answer to RQ3(b):] Semi-supervised label assignment significantly aids in evaluation. Grouping by exit codes alone achieves 98.7\% accuracy for binary bug detection. For granular root cause analysis, K-means clustering is required to separate distinct failure modes, improving propagation accuracy by 11.81 percentage points to 89.15\%. \end{tcolorbox} \end{center}

\newpage \section{Threats to Validity}
\label{sec:threats_to_validity}

Several threats to validity may affect the findings and broader interpretation
of this study. We discuss the potential threats to validity below.

\begin{itemize}
\item Internal Validity:

\begin{itemize}
    \item \textbf{Data Integrity:} The study relies on historical bug data and
          code output logs for benchmarking \textsc{SmartOracle}.
          Consequently, any mislabeling or inaccuracies in these foundational
          datasets could bias the recall metrics. We mitigated this by
          cross-referencing historical bugs with their upstream issue trackers
          to verify validity.
    
    \item \textbf{Environmental Fidelity:} A potential risk exists if the
          experimental environment (e.g., JavaScript engine build versions or
          OS configurations) does not precisely match the conditions under
          which historical bugs were originally reported. We attempted to
          replicate original environments using Docker containers, but minor
          discrepancies could alter observed runtime behaviors.
\end{itemize}

\item \textbf{External Validity:}

\begin{itemize}
    \item \textit{Domain Specificity:} The evaluation focuses exclusively on
          major JavaScript engines. While JavaScript represents a highly
          complex, dynamic execution environment, the results may not fully
          generalize to other programming languages (e.g., static languages
          like C/C++ or Rust) or different fuzzing targets (e.g., network
          protocols).
    
    \item \textit{Bug Distribution:} Differential fuzzing exhibits diverse
          characteristics depending on specification complexity. The findings
          derived from JavaScript's specific distribution of undefined
          behaviors may not be directly transferable to systems with stricter
          specification compliance or different error-handling paradigms.
\end{itemize}

\item \textbf{Construct Validity:}

\begin{itemize}
    \item \textit{Subjectivity of Oracle Ground Truth:}
          \textsc{SmartOracle}'s accuracy is measured against labeled bugs;
          however, determining whether a differential finding is a ``true bug''
          often involves subjective decisions regarding specification
          conformance and severity. These decisions may not be universally
          agreed upon, particularly for edge cases in the ECMAScript
          specification.
    
    \item \textit{Labeler Bias:} The false positive rate and label propagation
          accuracy are estimated via manual review of clusters. This introduces
          potential reviewer bias, especially if the status of a finding
          involves subtle nuances in specification. To mitigate this, we
          employed a strict labeling protocol based on consensus with the
          specification documentation.
\end{itemize}

\item \textbf{Conclusion Validity:}

\begin{itemize}
    \item \textit{Closed-Source Model Stochasticity:} The study utilizes
          commercial, closed-source Large Language Models (LLMs). These models
          are non-deterministic by nature and subject to opaque backend updates
          (model drift) by the provider. While we utilized low-temperature
          settings to maximize consistency, the exact reasoning and performance
          metrics may vary if the underlying model weights are modified, posing
          a threat to strict reproducibility.
    
    \item \textit{Algorithmic Sensitivity:} The observed reduction in manual
          triage effort stems from clustering and semi-supervised approaches.
          The effectiveness of these methods may depend on specific
          characteristics of the dataset. Different noise levels or fuzzing
          input sources could alter the optimal hyperparameters for the
          clustering algorithms, though our use of the unsupervised Elbow
          Method aims to mitigate this.
\end{itemize}
\end{itemize}

\section{Discussion}

\subsection{Semi-Supervised Learning: Important?}

A significant methodological contribution of this work is validating
semi-supervised label propagation for oracle evaluation. The results in
Figure \ref{sec:rq3b_results} demonstrate that clustering-based label assignment
achieves 98.7\% accuracy for binary bug classification and 89.15\% for
granular root cause analysis, addressing a critical bottleneck: constructing
ground-truth negative sets without exhaustive manual labeling. Exit code
patterns alone suffice for binary classification, enabling rapid dataset
expansion. For root cause analysis, K-means clustering on TF-IDF features
improves accuracy by 11.81 percentage points, with dramatic gains for
heterogeneous patterns (pattern\_7\_3\_0\_1: 28.57\% to 92.86\%). The
unsupervised Elbow Method automatically adapts to pattern structure, enabling
active learning workflows where oracle development transforms from exhaustive
labeling into efficient guided curation. We assert that (a) this use of
semi-supervised learning is a significant contribution of this paper; (b)
semi-supervised learning will be an important aspect in future fuzzing oracle
research.

\subsection{  Architecture Beats Scale?}

Beyond demonstrating effectiveness, understanding why
\textsc{SmartOracle} outperforms larger models reveals fundamental insights
about LLM deployment for specialized technical tasks. The 16-point recall
improvement (0.84 vs 0.68) despite fewer parameters validates that specialized
orchestration overcomes raw model scale. The 4x speedup and 10x cost reduction
stem from conditional execution: only relevant sub-agents activate for each
finding. Critically, combined sub-agent prompts exceed baseline length,
proving efficiency reflects smart execution paths, not prompt brevity. This
architectural pattern extends beyond differential fuzzing:
Xu et al. \cite{xu2025hallucinationconsensusmultiagentllms} achieve a 21.1 percentage point improvement in test oracle
correctness through multi-agent panel discussion, while He et al. \cite{he2025llm}
demonstrate scalable frameworks across the software lifecycle, suggesting
agentic decomposition is fundamental for LLM application to complex technical
tasks.

\subsection{Domain-Specific Challenges}

JavaScript's maturity as extensively studied software creates a specific
challenge profile: fuzzer output is dominated by well-understood benign
discrepancies (shell function variations, legacy features like
\texttt{\_\_proto\_\_}, web compatibility exceptions) rather than novel
specification violations. The signal-to-noise ratio can exceed 100:1 in
production, making false positive suppression critical. Currently,
\textsc{SmartOracle} requires manual specification of these noise patterns, a
limitation we address in future work through automated pattern learning
approaches.

\begin{figure}[h]
    \centering
    \includegraphics[width=\linewidth]{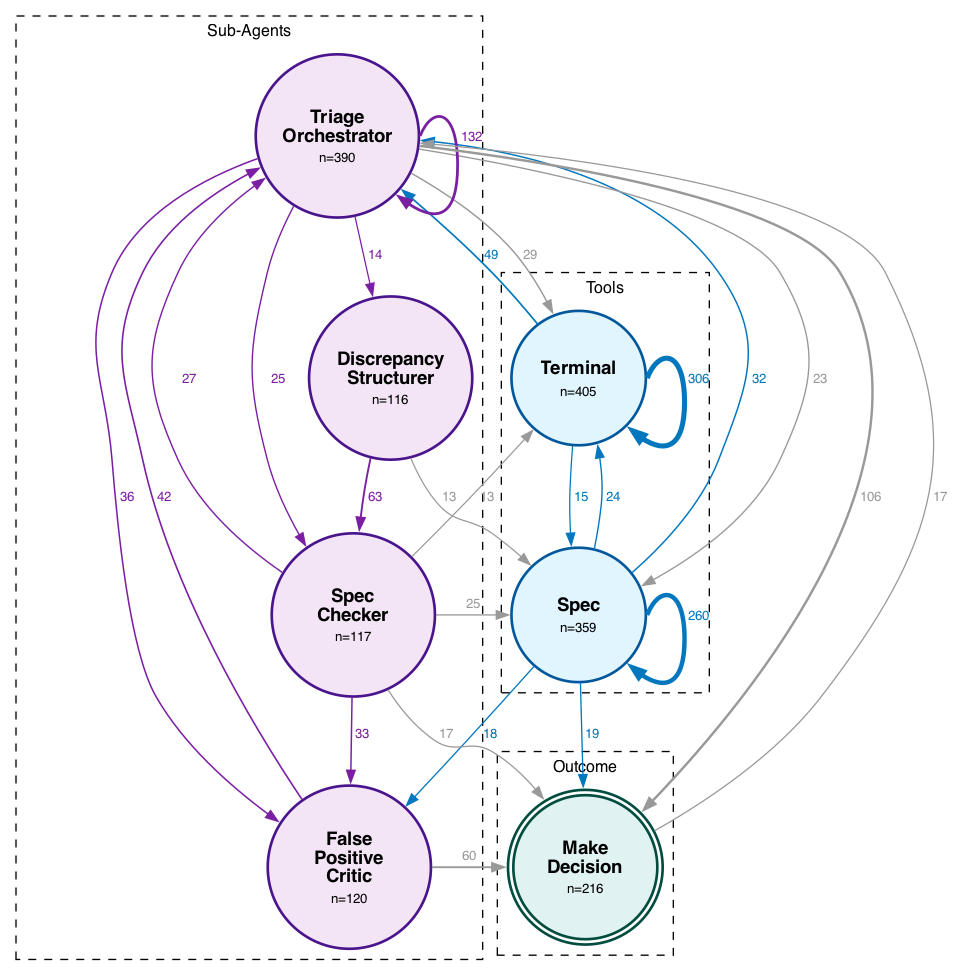}
    \caption{Agent and tool interaction flows in the triage process
             (234 logged reasoning steps)}
    \label{fig:reasoning_flow}
\end{figure}

\begin{figure}[h]
  \centering
  \begin{minipage}{0.48\textwidth}
    \centering
    \includegraphics[width=\linewidth]{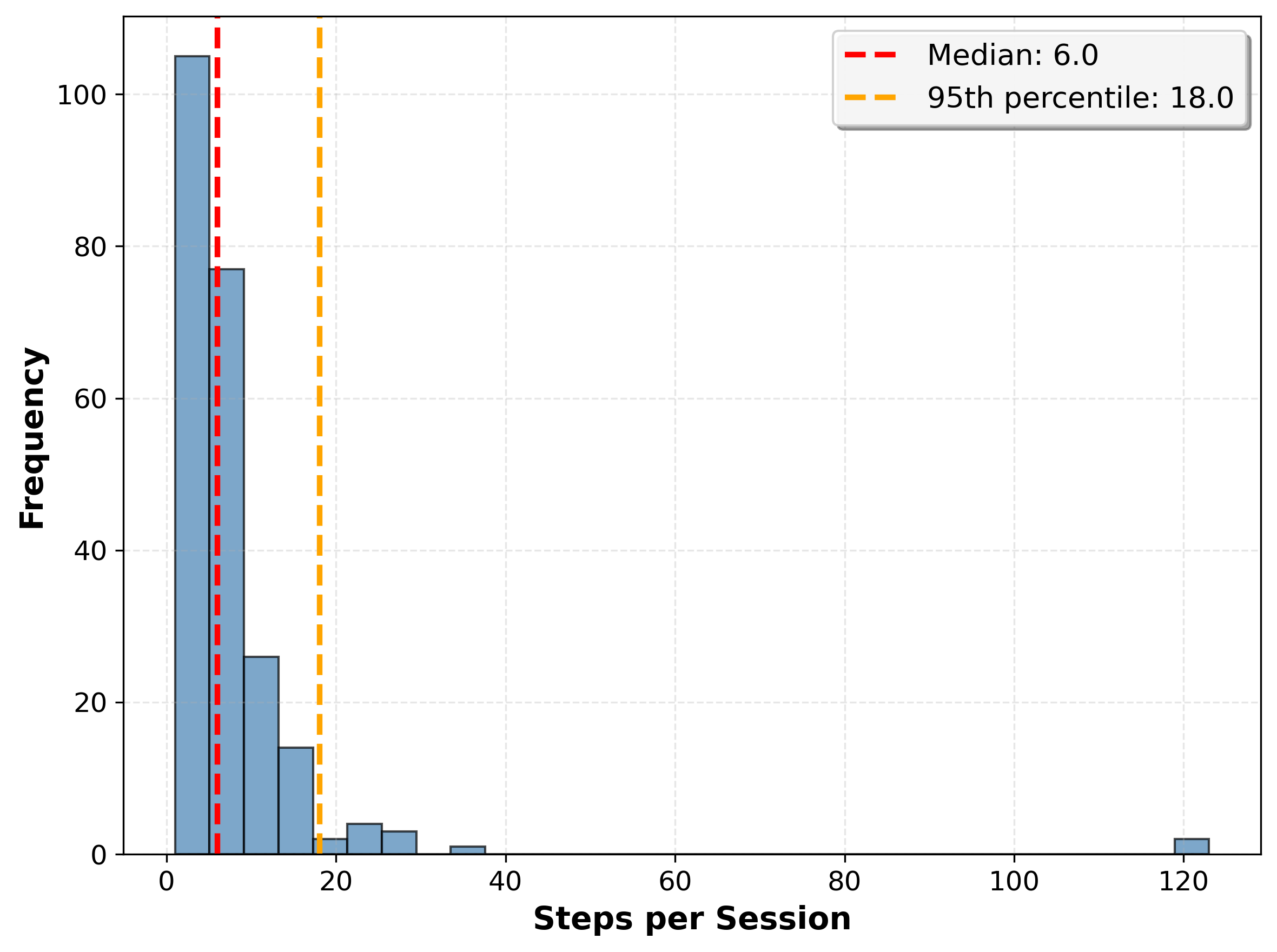}
    \caption{Number of reasoning steps before decision}
    \label{fig:step_count_distribution}
  \end{minipage}
  \hfill
  \begin{minipage}{0.48\textwidth}
    \centering
    \includegraphics[width=\linewidth]{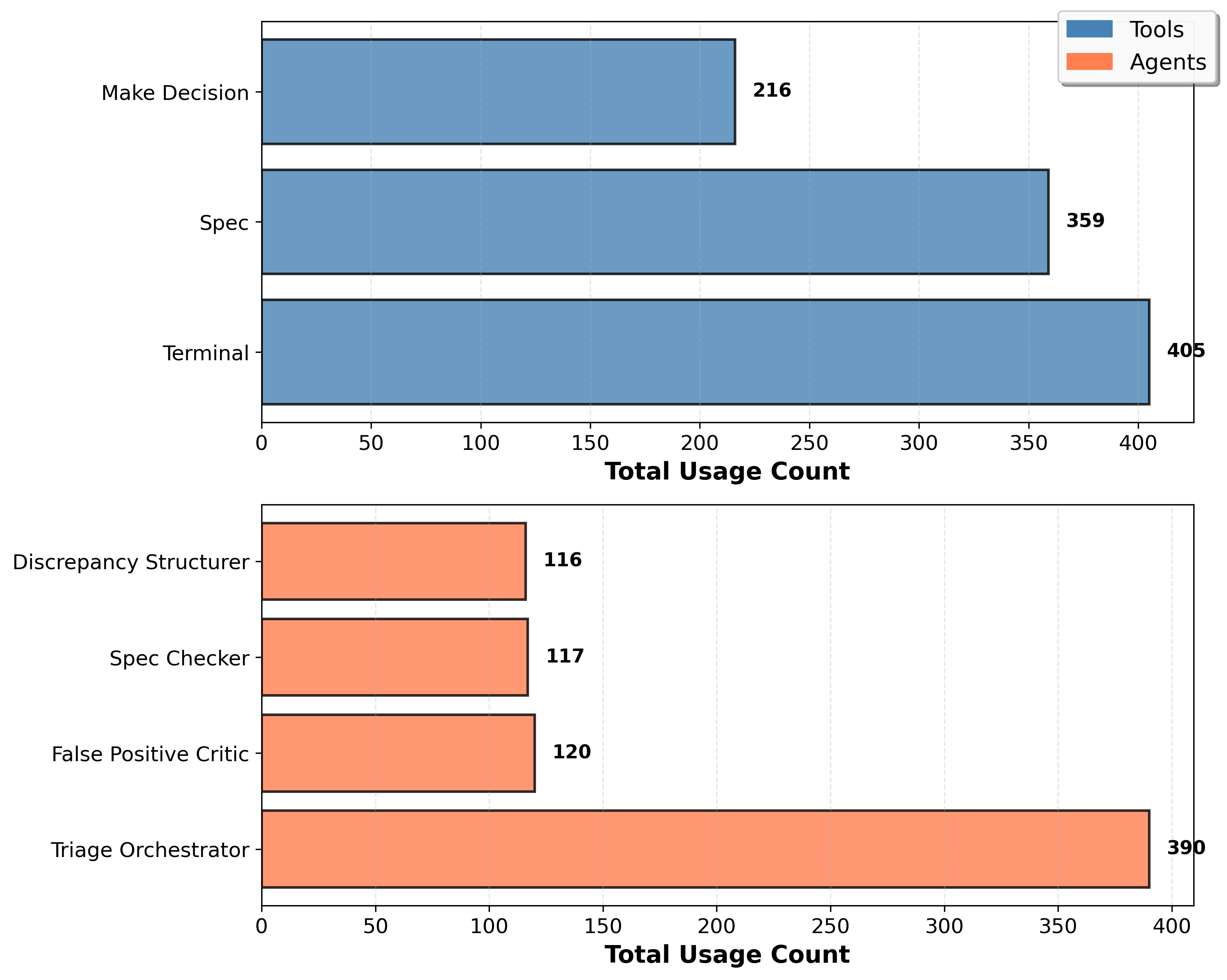}
    \caption{Distribution of tool and sub-agent usage}
    \label{fig:usage_heatmap}
  \end{minipage}
\end{figure}

\subsection{Tools Over Agents?}

To understand how \textsc{SmartOracle} operationalizes its multi-agent
architecture, we examine 234 logged reasoning traces. Figure~\ref{fig:reasoning_flow}
visualizes the complete interaction network between sub-agents, tools, and
decision outcomes. The orchestrator (n=390 calls, 132 self-loops) coordinates
all workflows, while tool interactions dominate: terminal (n=405) and spec (n=359) far exceed individual sub-agent invocations (each n=116-120).

Figure~\ref{fig:step_count_distribution} reveals the decision process is
remarkably efficient: median 6 steps, with 95th percentile at 18 steps.
Outliers reaching 120 steps represent reasoning loops where the system becomes
trapped; a hard limit prevents indefinite execution. Approximately 5\% of
cases encounter timeouts or parsing errors, an engineering issue addressable
through better agent frameworks.

Figure~\ref{fig:usage_heatmap} shows evidence gathering cycles between
terminal and spec tools through high-frequency terminal→terminal and spec→spec
transitions, reflecting iterative cross-engine testing and specification
queries. When primary tools prove inconclusive, specialized sub-agents
activate: the discrepancy structurer, spec checker, and false positive critic
(which notably precedes most SKIP decisions).

This pattern has design implications: rather than proliferating agent
communities, effort may yield greater returns by elaborating rich,
domain-specific tool interfaces that agents can leverage effectively. The data
suggests tool quality matters more than agent quantity for this task.

\section{Conclusion}

Differential fuzzing oracles face formidable challenges: high volume of
findings, false positives, and costly manual labeling. SmartOracle addresses
these through a scalable, multi-agent workflow combining semantic analysis,
specification-awareness, and automated noise suppression.

\textsc{SmartOracle} achieves 0.84 recall with 18\% false positive rate,
outperforming sequential LRM baselines with 4x speedup and 10x cost reduction.
Deployment discovered and confirmed specification-level bugs in actively
maintained JavaScript runtimes, including one fixed upstream (GraalJS \#931).
Semi-supervised label propagation achieves 98.7\% accuracy for binary
classification and 89.15\% for root cause analysis, enabling low-cost
construction of evaluation datasets. Smaller models with specialized
orchestration outperform larger monolithic models through conditional
execution and domain-specific tool interfaces.

These results validate structured agentic pipelines, supplemented by
semi-supervised techniques, as robust foundations for next-generation
differential oracles. Such systems significantly reduce manual review,
accelerate feedback cycles, and raise standards for automated software quality
assurance in rapidly evolving ecosystems.

\section{Future Work}

\textsc{SmartOracle} showcases the promise of agentic, LLM-driven oracles
paired with semi-supervised label propagation, but substantial research
avenues remain. Although validated within JavaScript engines, our approach is
likely transferable to other complex software systems, such as Python
interpreters, database backends, compiler toolchains, and network protocol
implementations. Each new domain introduces specification ambiguities and
legacy exceptions that may require adapting agent prompts, clustering
features, and underlying knowledge sources. Systematic studies exploring these
adaptations will be necessary to establish generalizability and uncover
unanticipated challenges beyond JavaScript. Statically-typed language runtimes
(C++, Rust, Java) where specification compliance may be more rigid but
implementation-defined behaviors still create differential oracle challenges
represent particularly promising targets. Similarly, databases with complex
query optimizers or distributed systems with subtle consistency models
represent domains where semantic analysis of differential behaviors could
yield substantial debugging value.

Improving the handling of noisy or non-standard findings remains critical.
Currently, our system requires specifying, in natural language, particular
shell artifacts, legacy behaviors, or web compatibility exceptions for the
agent to filter. This manual pattern definition constitutes a bottleneck and
limits scalability. Future work should develop targeted, automated mechanisms
for recognizing and learning these patterns, leveraging direct feedback,
cross-engine metadata, or continual learning. One promising approach involves
mining historical bug reports and their resolutions to automatically extract
patterns of confirmed false positives. Active learning techniques could
solicit human feedback on uncertain classifications, using this feedback to
refine filtering rules without requiring exhaustive upfront enumeration of all
noise sources.

While empirical results demonstrate \textsc{SmartOracle}'s effectiveness, the
theoretical foundations of agentic oracle systems remain underexplored. Future
research should investigate formal guarantees on recall bounds, false positive
rates, and convergence properties of multi-agent deliberation. Additionally, the stochastic nature
of LLM outputs necessitates robustness analysis and development of
probabilistic frameworks that model the reliability of multi-agent consensus
decisions.

The dominance of tool interactions in our reasoning traces suggests that
expanding tool capabilities could yield substantial gains. Future work should
explore symbolic execution integration to systematically explore divergent
code paths, differential static analysis comparing abstract syntax trees or
control flow graphs across engines, integration with historical bug databases
to provide precedent-based reasoning, and tighter coupling with automated
minimization techniques beyond our current test case minimizer sub-agent.
Transitioning from research prototype to production-ready oracle requires
addressing engineering challenges including the observed 5\% error rate due to
timeouts and parsing failures, streaming interfaces for real-time triage as
fuzzing campaigns generate findings, incremental learning to update agent
knowledge without full retraining, multi-tenancy for concurrent campaigns, and
comprehensive audit trails for debugging oracle decisions. Finally, advances
in LLM architecture, agent design, and active learning open opportunities for
more efficient, reliable, and explainable triage systems. Further refinements
to semi-supervised labeling could make data curation even more accurate and
cost-effective, while standardized benchmarking, robust agent traceability,
and practical integration with operational CI pipelines will ensure findings
remain relevant as both oracles and software evolve rapidly within the field.

\bibliographystyle{ACM-Reference-Format}
\bibliography{sample-base}

\appendix









\end{document}